\documentstyle[12pt]{article}

\def\thebibliography#1{\section*{{\hfill{References}\hfill}
\markright{References}}
\list
  {[\arabic{enumi}]}{\settowidth\labelwidth{[#1]}\leftmargin\labelwidth
\advance\leftmargin\labelsep\usecounter{enumi}}\sloppy\sfcode`\.=1000\relax}

\def\appendix{\section*{{\hfill{Appendix}\hfill}
\markright{Appendix}}\indent\noindent
\renewcommand{\theequation}{A\arabic{equation}}}

\renewcommand{\thesection}{\arabic{section}.}
\renewcommand{\theequation}{\thesection\arabic{equation}}

\def\sect{\setcounter{equation}{0}\section}

\newcommand{\bqn}{\begin{eqnarray}}
\newcommand{\eqn}{\end{eqnarray}}
\newcommand{\nn}{\nonumber}
\newcommand{\LL}{{\cal L}}
\newcommand{\bra}{\langle}
\newcommand{\ket}{\rangle}

\def\simle{\ \lower -2.5pt\hbox{$<$} \hskip-8pt \lower 2.5pt \hbox{$\sim$}\ }

\def\simge{\ \lower -2.5pt\hbox{$>$} \hskip-8pt \lower 2.5pt \hbox{$\sim$}\ }

\newcommand{\hfb}{\vspace{12pt}}

\title{\Large \bf
 DISORIENTED CHIRAL CONDENSATES:\\
A DYNAMICAL SIMULATION
IN THE (2+1)-DIMENSIONAL GROSS-NEVEU MODEL \footnote{Partially
supported by the Swiss National Foundation}}

\author{
\\
A.Barducci, L.Caiani, R.Casalbuoni, M.Modugno, G.Pettini\\
{ \small Dipartimento di Fisica, Univ. di Firenze} \\
{\small I.N.F.N., Sezione di Firenze }\\[0.8cm]
R.Gatto\\
{\small D\'epartement de Physique Th\'eorique, Univ. de Gen\`eve}
}\date{}

\begin{document}
\maketitle
\vspace{5cm}
\centerline{UGVA-DPT 1995/07-898}
\thispagestyle{empty}
\newpage

\centerline{\large \bf Abstract}\hfb

\begin{quotation}
We simulate the formation and growth of disoriented chiral condensate
(DCC) regions which
follow the expansion of a high energy density region
into the ``cold'' vacuum.
The numerical study is based
on the one-loop effective potential for the massive 2+1-dimensional
Gross-Neveu model. We pay attention to the setting of the initial conditions
and to determining which parameters are relevant for
a strong amplification of the pion field. We find that the size of the
``hot'' source plays a significant role.
For large enough source radii, we observe strong correlation
phenomena, corresponding to the growth of large regions where the pion field
oscillates along a given direction.
We give our results in terms of the $\theta$ angle which defines
the DCC disorientation, of the other $O(4)$ angles distributions, of
the local ratios $R_{a}=\pi_{a}^{2}/{\vec{\pi}}^2$ $(a=1,2,3)$, and of the
energies associated with the fields at representative times.
\end{quotation}

\thispagestyle{empty}
\newpage

\sect{Introduction}
\indent\noindent

It is well known that QCD has an approximate
$SU(2)_{L}$ $\otimes$ $SU(2)_{R}$ global chiral symmetry and that
the pions are the pseudo-goldstone bosons associated
to the spontaneous breaking of the symmetry down to $SU(2)_{L+R}$.
On the other hand, asymptotic freedom implies that there should
be qualitative changes in the properties of hadronic matter as the
temperature and/or the baryonic density are increased. Thus we expect a
colour deconfining phase transition. Also another
phenomenon is expected, which is just the restoration of chiral symmetry.
There are indications that these phase transitions occur at the same critical
temperature, presumably located around $150~MeV$ \cite{dec}.
Since such temperatures are supposed to be attainable in
high-energy collisions of heavy ions, much effort has already
been done to identify possible experimental signatures
of the phase transition, primarily of the deconfinement\cite{qcd}.

One very interesting proposal on
how to decide whether the restoration of chiral symmetry
was reached or not during a heavy ion collision was suggested during the last
few years \cite{ans,bjo,bk,kt,rw}.
The basic mechanism is the following:
closely above  the chiral critical temperature
$T_{c}$, chiral symmetry implies the equivalence of the different
orientations in the internal space.
The non-equilibrium expansion and consequent cooling of the plasma
makes however the chiral symmetry to break and the direction of the symmetry
breaking in the internal space does not necessarily have to be the same
as it originally was before the collision took place.
The small chiral breaking terms are presumably negligible at this stage.

To explain the possible physical consequences, the language of the linear
$\sigma$-model \cite{sigma} has been widely employed,
since the relevant part of the order parameter is a four-component
vector $\phi^{\alpha}\equiv(\sigma, \vec\pi)$,
 and the chiral transformations are
$O(4)$ rotations in the internal space.
In terms of these fields,
the usual zero-temperature vacuum is defined by
$\langle \sigma \rangle = f_{\pi}$ and $\langle \vec\pi \rangle = 0$.
Thus, the situation depicted above, indicates the possibility that before
the plasma approaches again equilibrium, in some region of space-time
the vacuum orientation can be different from the usual one, namely:
$\langle \sigma \rangle = f_{\pi} \cos \theta$ and
$\langle \vec\pi \rangle = f_{\pi}\vec n \sin \theta$,
where $\vec n$ is a some
unity vector in the internal $O(3)$ pion space.
Because of the fact that the pion mass is
small, it costs relatively little energy, of the order of
$f_{\pi}^{2} m_{\pi}^{2}$ per unit volume, to disorient the
vacuum. Macroscopic
domains containing a quasi-classical pion field can appear
as the temperature drops below $T_{c}$ \cite{kt,rw}.

Such metastable domains will eventually
disappear as the system evolves towards the true vacuum,
mainly via $O(3)$-coherent emission of pions.
The formation of large domains could lead to events
with an unusually large number of either neutral $\pi^{0}$ or charged
$\pi^{\pm}$ pions (Centauro and anti-Centauro events \cite{centauro}).

The scenario that we have summarized is referred to in literature as
Disoriented Chiral Condensates (DCC) \cite{bjo,bk,kt}.

A specific prediction for the distributions of the fraction of neutral
pions emerging from each domain
can be obtained from the following
semiclassical arguments (see e.g. \cite{kt,rw} and
refs. therein).  Let us consider a single (ideal) domain in which the
pion field is oscillating along a fixed direction
$\vec n$. Classically, pion production from this domain is
proportional to the square of the field strength (think of
non-relativistic coherent states),
so that the fraction of the neutral pion is $R=\pi_{0}^2/{\vec\pi}^{2}$.
Since all orientations of $\vec\pi$ are a priori equally likely, the
probability of finding a fraction R of neutral pions is
$P(R)=1/(2\sqrt{R})$ (see for instance \cite{rw}).
In the case of events dominated by a
single large domain, this feature would thus reflect on the distribution
probability\footnote{ Probability on a event-by-event basis, i.e. a
distribution over many events.}  of the observed fraction of neutral
pions, whose shape strongly deviates from the binomial distribution
of typical hadronic collisions.
We remark that the number of soft pions produced at high energies
in such collisions could be quite large, which makes plausible the
validity of semiclassical arguments. In fact, it seems possible
that at high energy density the produced pions can create a certain
coherent state,
i.e. a classical pion field \cite{ans}.\hfb

In order to study these problems one should consider, strictly speaking, the
real hadron world (quarks and gluons) with all its
complexities. Unfortunately, the detailed dynamics of heavy ion physics is far
beyond what can be calculated directly from QCD.
Therefore most discussions
of DCC are necessarily based on simplified models which
can describe the low energy behaviour of QCD.

Actually in recent literature, the effective Lagrangian of the $O(4)$ linear
$\sigma$-model has been used to show, by numerical
simulations, the amplification of soft pion modes following a
quench from high to low temperatures \cite{rw}.
Other numerical studies have incorporated the
feature of the expansion of the plasma by starting from a spatially
non uniform energy distribution, and have progressed also
in discussing the role of the fluctuations \cite{ahw}.

Other interesting studies on the subject can be found in ref.
\cite{glb,ggp,gm}.

In this work we present a numerical study, performed
on the basis of the one-loop effective potential
of the Gross-Neveu model in 2+1 dimensions, which we believe can
retain in these connections some aspects of real QCD \cite{gn}.

Let us briefly comment on the choice of this model. Though less
phy\-si\-cal, a two-space-di\-men\-sio\-nal model allows for faster computer
simulations, thus opening the possibility of a more thorough analysis.
Some main features of the Gross-Neveu model in 2+1 dimensions
are known, even if in an equilibrium context and in the mean field
approximation \cite{rwp}.  Thus, in principle, one could compare some
equilibrium property (for instance, in order to better discuss the
initial conditions of a quench), as derived from a microcanonical
simulation and from the canonical-ensemble study of ref.\cite{rwp}.
Furthermore, the one-loop zero temperature effective potential of
the model, expressed in terms of the appropriate scalar and pseudoscalar
composite fields, comes directly from the integration over fundamental
fermion fields of the underlying theory, and has a
very simple expression sharing the principal features of the
$\sigma$-model.  Incidentally, the form of the potential
 raises the problem of better exploring
the effects of different kind of non-linearities.\hfb

Coming back to the phenomenology of DCC, we simulate the amplification
of the pion field following a quench, i.e. a sudden transition from
high to low temperature (from $T$ above $T_{c}$ to $T$ below $T_{c}$).
In our case the quench is realized by means of the expansion of a
high-energy-density region (in which the field fluctuates around the
maximum of the double-well potential), into the cold vacuum
surrounding it (in which the field performs small-amplitudes
fluctuations around the absolute minimum of the potential)
\cite{ahw}.\hfb

In such a simulation, the only ``thermal'' character of the system
lies in the initial conditions. Consistently, we pay attention to
controlling the role of the parameters which describe our initial
conditions, such as the energy density, the source radius etc., and to
build up a simulation scheme suited to give the results in terms of
physically measurable quantities, such as the energies of the outgoing
pions and their number.  For the time being, we do not worry too
much about a physically quantitative description which
will be hopefully gained only by
repeating the study in a more realistic 3+1 dimensional model; here we
concentrate rather on the observation of the basic mechanisms leading
to the amplification of the pion field, as a guidebook
to other simulations.\hfb

In Sect.2 we
summarize the basic features of the Gross-Neveu model and
of the effective action we use. In Sect.3
we present and discuss in more detail the numerical
simulation.
This section is divided in subsections, containing
a discussion of the initial conditions, the analysis
of the different quantities we have measured, and the
sensitivity of the results to changes in the parameters.

In Sect.4 we summarize the results in the conclusions.\hfb

\newpage
\sect{The Model}
\indent\noindent

In order to study the possible formation of DCC, and its
phenomenological consequences, we have carried
out numerical simulations for the (2+1)-dimensional
massive Gross-Neveu model \cite{gn,rwp}.

The model is defined by a four-fermion interaction Lagrangian density
\bqn
{\cal L}={\bar\psi}\left(i{\hat\partial}-M\right)\psi
+{g^2\over 2}\Big[({\bar\psi}\psi)^2 - ({\bar\psi}{\vec\tau}\gamma_{5}\psi)^2
\Big]\label{lgn}
\eqn
where $M$ is the bare fermion mass,
$\tau_i$ are the matrices of the fundamental representation
 of $SU(N_f )$, and $\psi$ consist of $N$ separate $N_f$-dimensional
representation of $SU(N_f )$. In the following we take $N_f =2$.

For $M=0$ the model has an exact chiral
$SU(2)_{L}$ $\otimes$ $SU(2)_{R}$ symmetry
which is dynamically broken to $SU(2)_{V}$. It is renormalizable
in the $1/N$ expansion \cite{rwp,ren}.

The four-fermion interaction can be conveniently studied in the $1/N$ expansion
by introducing two auxiliary fields $(\sigma,\vec\pi)$ which satisfy
the classical equations of motion
\bqn
\sigma &=&g{\bar\psi}\psi-{M\over g}\nn\\
{\vec\pi}&=&g{\bar\psi}i\gamma_{5}{\vec\tau}\psi
\eqn
Then the Lagrangian becomes
\bqn
{\cal L}={\bar\psi}i{\hat\partial}\psi - {1\over 2}(\sigma^2+{\vec\pi}^2)
+g{\bar\psi}(\sigma+i\gamma_{5}{\vec\tau}\cdot{\vec\pi})\psi
-M{\sigma\over g}
\eqn
and by integrating over the fermion fields one can study the effective
action as a functional of $(\sigma,{\vec\pi})$.

The effective Lagrangian obtained by this procedure is
\bqn
{\cal {L}}_{eff} = {1\over 2c_{0}^2} \left(\partial_{\mu}\sigma
\partial^{\mu}\sigma + \partial_{\mu}{\vec\pi}\cdot\partial^{\mu}{\vec\pi}
\right)-
{\cal {V}}_{eff}
\label{leff}\eqn
where derivative terms have been kept only up to second order.
The one-loop effective potential ${\cal {V}}_{eff}$
is the sum of a chirally symmetric part \cite{rwp}
and a breaking term proportional to $\sigma$
\bqn{\cal {V}}_{eff}(\sigma^2+{\vec\pi}^2,\sigma)=
{N_f Ng^3 \over 2\pi}\left({(\sigma^2+{\vec\pi}^2)^{3/2}\over 3} -
{m_{0}\over g}{(\sigma^2+{\vec\pi}^2)\over 2}\right)
+ {M\sigma\over g}
\eqn
$m_{0}$ being the dynamical mass
acquired by the fermions due to the spontaneous breaking
of chiral symmetry \footnote{The finite temperature extension
of this effective potential for $M=0$ in the Canonical Ensemble
has been calculated and discussed in ref.\cite{rwp}. The mean
field calculation shows the existence of a second order phase
transition at $T_{c}=m_{0}/2\ln 2$, whilst by taking into account
the field fluctuations, this transition turns out to be of the
Kosterlitz-Thouless type (see refs.\cite{rwp,bar} and references therein).}

Clearly, for dimensional reasons, the $\sigma$ and $\vec\pi$ fields
cannot be identified with the canonical ones.
The $c_{0}$ factor in eq.(\ref{leff}),
which rescales the fields to the canonical ones
$(\varphi_{\sigma},{\vec\varphi}_{\pi})
{}~\equiv~ (\sigma,{\vec\pi})/c_{0}$,
can be determined
by requiring that the effective potential ${\cal {V}}_{eff}$
gives the correct $\sigma$-mass,
$m^2_\sigma = 4m_{0}^2$ \cite{rwp,bar}. The value we get
is $c_{0}^2=8\pi m_{0}/(N_f Ng^2)$.

After some algebra, the effective Lagrangian can be written as
\bqn
{8\pi\over N_{f}Nm_{0}^3}\LL_{eff}={1\over2 m_{0}^2 }\partial_{\mu}
\phi_{\alpha}\partial^{\mu}\phi_{\alpha} -U(\phi_{\alpha})
\label{lad}\eqn
\bqn
U(\phi_{\alpha})={4\over 3}(\phi_{\alpha}\phi^{\alpha})^{3/2}
{}~-~2\phi_{\alpha}\phi^{\alpha}~+~\lambda\phi^{\alpha}\delta_{\alpha 1}
\label{pot}\eqn
where we have defined
\bqn
\phi_{\alpha}\equiv{g\over m_{0}}(\sigma,\vec\pi)\quad\quad
\lambda\equiv\frac{8\pi M}{N_f Ng^2 m_{0}^2}
\label{adfield}\eqn
In Fig.\ref{potential} we plot the effective potential
(\ref{pot}) $U(\phi_{1},{\vec 0})-U(v,\vec{0})$.
Its extrema are located at
\bqn
v~&=&-{1\over 2}~\left(1+\sqrt{1+\lambda}\right);
{}~~~~~~~~~~~~~~~~~~abs.~min.\nn\\
\phi_{0}~&=&~{1\over 2}~\left(1-\sqrt{1-\lambda}\right);
{}~~~~~~~~~~~~~~~~~~~rel.~max.\label{extrema}\\
v_{+}~&=&~{1\over 2}~\left(1+\sqrt{1-\lambda}\right);
{}~~~~~~~~~~~~~~~~~~~rel.~min.\nn
\eqn
We have set the parameters to $m_{0}=300 MeV$ and $\lambda=1/4$,
leading to $m_{\sigma}=635MeV$ and $m_{\pi}=145MeV$.

For the nu\-me\-ri\-cal si\-mu\-la\-tion we employ
di\-men\-sion\-less
va\-ria\-bles
\bqn
({\vec\xi},\tau)=(m_{0}{\vec x},m_{0}t)\label{xitau}
\eqn
and, from eq.(\ref{lad}),  the Hamiltonian
\bqn
H~\equiv ~{8\pi\over N_f Nm_0^3}{\cal{H}}_{eff}~=~
\sum_{\alpha=1}^{4}{{\dot{\phi}}_{\alpha}^{2}\over 2}~+~
\sum_{\alpha=1}^{4}{(\nabla {\phi}_{\alpha})^2\over 2}~+~U(\phi_{\alpha})
-U(v,\vec{0})
\label{ham}
\eqn
where the derivatives are now taken with respect to
$({\vec\xi},\tau)$,
 and the vacuum energy has been subtracted.

\newpage
\sect{Dynamical Simulation}

\subsection{Formulation}
\indent\noindent

Our aim is to simulate, within the model, the dynamics of the
expansion of a high energy density region in a low energy
density environment.
The first region should simulate the hot fireball
after a central collision and the consequent compression,
whereas the second region is the usual vacuum.

The central assumption is that,
before the fireball begins to expand into the vacuum,
it can be viewed as an isolated system close to
thermal equilibrium near the critical
temperature\footnote{As far as the average
value of some observables is concerned.},
in a region of linear size of the order of some $fm$.

In view of the initial conditions we want to represent,
the configuration of the $\sigma$ field must be
chosen in a way that the mean value $\bra \sigma\ket$ is close to
zero\footnote{It is not strictly vanishing due to the
presence
of an explicit symmetry-breaking term in the Lagrangian (\ref{lgn})}.
This is achieved, in the central region,
by assigning to the $\sigma$-field a constant value
corresponding to the relative maximum of the potential
(\ref{pot}). The same we do for the $\vec\pi$ field,
with the difference that the initial configuration is now
strictly zero.

To start the dynamics, we assign initial velocities
according to a random gaussian distribution with a given variance,
the same for the $\sigma$ and $\vec\pi$ fields, as they have to be
degenerate in the source (apart for the symmetry breaking term).
We remark that,
since the mean value of the initial velocities is taken as vanishing,
to assign the velocity variance corresponds to determine the
first term of the Hamiltonian of eq.(\ref{ham}).

Out of the central region, the $\sigma$-field
should smoothly connect with the cold vacuum where the
velocity fluctuations are small, due to the smallness of the
energy density.
Within the source, the $\vec\pi$-field initial
amplitude is set constant and equal to zero.

The intermediate region, which connects the ``hot" vacuum to the
``cold" one, is the most energetic and the energy is carried
almost entirely by the $\sigma$-field.
In fact the gradient term of the Hamiltonian (\ref{ham})
grows significantly due to the
spatial variation of $\sigma$ from the edge
of the central region to the vacuum outside.
Such a growth overcomes the decrease of potential energy
if the variation of the field is steep enough in the connecting region.

Finally, in the third region, the ``cold" vacuum is represented
by putting the $\sigma$ at the absolute minimum of the potential
(\ref{pot}) and the pion field to zero, and giving to them
a nearly vanishing velocity variance.

The successive evolution is governed by the Hamiltonian
of eq.(\ref{ham}) and the
dynamics soon forces the system to have certain (fixed) ratios
between the three terms in the r.h.s. of eq.(\ref{ham}),
with the velocity fluctuations inducing fluctuations also in the
field amplitudes.\hfb

We note that a nearly
constant initial field amplitude provides an easy way to fix the
energy at the beginning, since it decouples the Hamiltonian and makes
then easier to keep the various terms under control: the successive
dynamical evolution then produces a non-constant amplitude, but the
energy remains fixed.  Moreover, since we are to
approximate a smooth field on the lattice, the field values must not
vary too much within a lattice spacing, otherwise the field would not
be differentiable in the continuum limit and the derivative terms
would diverge. Thus we find incorrect to choose an initial field
configuration distributed randomly over a length scale of the order of
the lattice spacing, unless of nearly vanishing amplitudes. For
the same reason the initial velocity variance
(and thus the initial energy) cannot grow indefinitely in this scheme.

Let us remark that the gradient term is
related to the fluctuations of the field, but it
contains more information since it depends also on their spatial
distribution.\hfb

As far as the numerical simulation is concerned, we
solve the classical equations of motion
for the dynamics of the fields $\phi_{\alpha}(x,y,t)$ by using
laboratory space-time coordinates. The Hamilton equations
are integrated by means of a second order bilateral
symplectic algorithm (for the details of the method see
ref. \cite{case}, whereas a sketch of the method is in the Appendix)
on a $200\times200$ square lattice, with a
time integration step $dt=a/10$ where $a$ is the
lattice spacing, which we fix to $0.1 fm$
\footnote{We have verified that reducing the temporal step and
the lattice spacing does not change our results.
The simulation time is limited by the lattice dimensions, since
the expansion can be simulated until the energy of the source
reaches the lattice boundary.}.

\subsection{Initial conditions}
\indent\noindent

The standard initial conditions that we choose
are given by setting the following field configuration
\bqn
\phi(\vec{\xi},\tau = 0)=\left(
\begin{array}{c}
v + f(\rho)(\phi_{0} -v) \\
{\vec{0}}
\end{array}
\right)\label{initconf} \eqn where $(v,\vec0)$ and $(\phi_{0},\vec0)$
are respectively the absolute minimum and the relative maximum of
$U(\phi_{\alpha})$ in eq.(\ref{pot}), with values given by eqs.
(\ref{extrema}) ($\lambda=1/4$). The function $f(\rho)$
 is suited to
make $\phi_1=\phi_0$ in a central disk of radius ten sites, and to
connect to $\phi_{1}=v$ in about five more sites. This corresponds to have the
source containing the ``hot vacuum'' in a disk of radius $1~fm$,
separated from the ``cold vacuum'' by a ring of thickness of about
$0.5~fm$.

These features are realized by means of the following $f(\rho)$
\bqn f(\rho)=\left\{
\begin{array}{ll}
1&\rho<{\ell}\\
\exp\left\{\displaystyle{{1\over s}~\left[{1\over (\rho-{\ell})^2-w^2}~
+~{1\over w^2}\right]}\right\}
&{\ell}<\rho<\ell+w\\
0& \rho>\ell+w
\end{array}
\right.
\label{function}
\eqn
where $\ell=10$, $s=1/900$, $w=10$,
$\rho=r/a$ , $r$ being the distance from the lattice center.

As it is seen from eq.(\ref{function}), the tail of the
distribution extends over a radius of $2~fm$, nevertheless
$f(\rho)$ is sensibly different from zero only for $\rho a<1.5~fm$.
An aspect related to the length of the tail is discussed later on.

The initial velocities are given with $\langle {\dot\phi}_{\alpha}\rangle=0$
and variance
$\delta{\dot\phi}^{2}_{\alpha}\equiv\langle {\dot\phi}^{2}_{\alpha}\rangle$,
with values within $[0.01,1]$.
The velocity variance itself is then spatially modulated consistently
with the condition of having an high-energy-density and a
low-energy-density region. Furthermore, in order to produce
small field fluctuations in the ``cold'' vacuum, we modulate
$\delta{\dot{\phi}}^2_{\alpha}$ with the function $0.99~f(\rho)+0.01$.

In Figs.\ref{initialsigma}, \ref{initialpi} and \ref{initialen}
we plot respectively
the configurations on the lattice
of the $\sigma$-field, of the $\pi_{2}$-field,
and that of the energy density,
at $t=0.02 fm$ (one algorithm step).

The results of the simulation are shown in the following subsections.

\subsection{Disorientation of the chiral condensate}
\indent\noindent

To describe our numerical results, it is convenient to consider the
following parameterization of the $\phi$ vector in terms of the
$O(4)$ angles
\bqn
(\sigma,\pi_{1},\pi_{2},\pi_{3})=
\rho~({\rm{cos}}\theta,~{\rm{sin}}\theta~{\rm{cos}}\varphi,
{}~{\rm{sin}}\theta~{\rm{sin}}\varphi~{\rm{cos}}\eta,
{}~{\rm{sin}}\theta~{\rm{sin}}\varphi~{\rm{sin}}\eta)
\label{fiesta}\eqn
As anticipated in the introduction, the amplification of the pion field
following the quench becomes an important phenomenon
only when the $\theta$ angle undergoes
strong and coherent disorientations over large
enough regions, corresponding to a big transfer of energy to the pion field
itself. On the other hand, in order to have a visible effect on the
produced pions, it is crucial that this transfer of energy
takes place in a preferred direction in the internal $O(3)$ space, which
is signaled by the distribution of the $\varphi$ and $\eta$ angles.

These parameters are important expecially before the decoupling of the
pions fields, i.e. until when the $\theta$ angle performs small
amplitude oscillations around its equilibrium value $\theta_{eq}=\pi$.
Actually, it is evident
that after this time interval the more natural description is given directly
in terms of energy and number of the outgoing pions.

In Fig.\ref{thetaregions}, some $\theta$
configurations are shown for representative times. In this picture
we plot the $\theta$ angle on the internal
$32\times32$
square lattice centered in the source (which initially is a circle of radius
$\rho\simeq10$ sites), by dividing the range $[0,\pi]$ in five equally
spaced regions. Such regions are colored by using a gray scale,
ranging from black ($\theta\simeq0$) to white ($\theta\simeq\pi$).

According to Fig.\ref{thetaregions} we can
distinguish three different behaviours. At early times there is a
formation and growth of $\theta$ correlated regions inside the source.
Such regions tend to melt together in the bulk, while the outer
regions align to $\theta=\pi$.  This phenomenon ends up at about
$t=1fm$ (see Fig.\ref{thetaslope}),
when the whole source is aligned with the true vacuum.
Anyway, the system has not relaxed to the true vacuum yet. From the
center of the source, in fact, a collective motion starts, which
brings, at $t\simeq2.5fm$, the whole source to be at $\theta=0$.
Afterwards, the system begins to perform collective damped
oscillations along the bottom of the
potential well\footnote{ Notice,
for instance, that the system at $t=4fm$ and $t=6fm$ is aligned to
$\theta=\pi$, which is represented by the white color in the
pictures.}, while
transferring energy to the pion modes. In this event,
most of the energy is acquired by the $\pi_2$ field, and this happens
between $t\simeq4.5fm$ and $t\simeq7fm$. At later times,
$t\simge10 fm$, the system eventually decouples (see later).

In Fig.\ref{thetaslope}$a$ we plot the mean value
$\langle\theta (r)\rangle (t)$
(where the mean value is taken on concentric rings of
thickness $0.4 fm$) to point out the time scales
characterizing the different stages of the phenomenon of DCC.
In Fig.\ref{thetaslope}$b$ we further point out
(from Fig.\ref{thetaslope}$a$) the profiles
of $\langle\theta ({\bar {r}})\rangle (t)$ with
$0 <{\bar{r}}<0.4~fm$ and $0.8<{\bar{r}}<1.2~fm$.

In Fig.\ref{vectorfields} we show another representation which
is less advantageous as far as the growth of coherent regions
is concerned, but which better takes into account the small $\theta$
disorientations which are not visible in Fig.\ref{thetaregions}.
Thus in Fig.\ref{vectorfields} the $\theta$ field on the $(x,y)$
lattice is directly given by the orientations of the arrows.
These pictures are taken in coincidence of some maxima of
disorientation (see Fig.\ref{thetaslope}). As in
Fig.\ref{thetaregions}, it can be noticed that
the disorientation is spatially  incoherent
at $t=0.02fm$ (Fig.\ref{vectorfields}$a$), whereas
it has become coherent at $t=2fm$ and $t=5fm$
(Fig.\ref{vectorfields}$b$-$c$).
Finally, in Fig.\ref{vectorfields}$d$, at $t=9fm$,
the presence of concentric regions of different $\theta$
orientation is evident (see also Fig.\ref{thetaslope}$a$
by taking a profile at fixed $t$).

Figs.\ref{thetaregions},\ref{thetaslope} and \ref{vectorfields}
have all been
obtained with $\delta{\dot\phi}^{2}_{\alpha}=0.1$. The qualitative picture
does not change by varying the initial velocity variance (and thus the
initial energy), whereas the coherence of the phenomenon
reduces by raising $\delta{\dot\phi}^{2}_{\alpha}$ (in the interval
we have considered, i.e. $[0.01,1]$), due to the consequent
increase of the fields fluctuations.

\subsection{$O(3)$-coherence}
\indent\noindent

The dynamical evolution of the $\theta$ angle, that we have sketched
in the previous paragraph, reminds us a laser-like mechanism. In fact,
first most of the available energy is stored in an unstable ground
state, and later, when this state relaxes to the true vacuum, there is
a transfer of energy to the pion modes.

The important question now, is whether this energy is coherently
transferred to a single pion field over a large
region, or not. In the former case this could lead to a distribution
probability of the emitted pions which deviates from the usual
binomial one, as anticipated in the introduction (see for example
 \cite{kt,rw,ggp,gm}).

 To answer this question, and to better analyse the mechanism by which
 pion collective modes get enhanced, we have proceeded in two
 different ways. First we have looked, at fixed times, at the
 distributions of the representative points of the lattice sites in
 the $(\varphi,\eta)$ plane (see Fig.\ref{etaphi}) The existence of
 preferred directions in which the pion fields oscillate is signaled
 by the formations of clusters of points.

These pictures are particularly useful to signal the existence of
preferred oscillation directions, but they do not give any information
on the spatial correlation of such phenomenon. Thus, to improve
the description, and to analyse the formation and growth of spatially
correlated regions, we have considered the time evolution of the
ratios
\bqn
R_{a}=\pi_{a}^2/{\vec\pi}^{2}
\eqn
$(a=1,2,3)$ at each site of the lattice. These ratios define the orientation
of the oscillation direction with respect to the $a$-th field.

In the real situation, one is especially interested in the ratio
$R_{0}\equiv\pi_0^2 /\vec{\pi}^2$. Anyway, since in these
simulations we do not implement the electromagnetic coupling,
all the parameterization, and thus
each of the three ratios, are a priori equivalent.

Obviously none of these ratios is defined for vanishing pion fields
amplitudes and thus, in particular, they are not defined at the time
$t=0$, with our initial conditions.  But as the dynamics starts, the
pion field begins to perform oscillations in some direction of the
internal $O(3)$ space and thus the ratios $R_{a}$ are perfectly
defined. As the initial velocities are given randomly, the resulting
distribution is $P(R)=1/(2{\sqrt {R}})$, which is a consequence of the
initial equal likelihood of any $O(3)$ direction at each site.  To get
a pictorial representation of the $R_{a}$-regions, we have (see for
instance \cite{rw}) divided the interval of definition of $R$ in
three equally probable regions: black if $R\in
[0,1/9]$, grey if $R\in [1/9,4/9]$ and white if $R\in [4/9,1]$.  Since
the expected mean value is $\langle R\rangle=1/3$, black and white
regions represent regions of big deviations from the mean value $1/3$.
In Fig.\ref{R2}$a$ we show the spatial distribution of $R_{2}$
at the initial time $t=0.02fm$.  The equality of the areas and the
mean value $1/3$ are very well respected. Anyway, the ratios $R_{a}
(x,y)$ and the mean values $\bra R_{a} \ket$ are not conserved
quantities\footnote{ The conserved Noether currents are the isospin ones.},
 and
thus the equality of the areas has not to be preserved during the
evolution of the system. This is strictly related to what we are
looking for: the formation of few large regions where the pion field
oscillates coherently around a particular direction.

In the other pictures of Fig.\ref{R2}, we show
the configuration of the ratio $R_{2}$ over the lattice at representative
times. We see that, as already mentioned, at $t\simeq4fm$ a
pion collective motion begins. This leads to the formation of large $R$
correlated regions, where the pion field points to a unique direction
in the $O(3)$ space  (see Fig.\ref{etaphi}).
The plots in Figs.\ref{etaphi}, \ref{R2} refer to the same event of
Figs.\ref{thetaregions}, \ref{vectorfields} and thus are for
$\delta{\dot\phi}^{2}_{\alpha}=0.1$. The same considerations done at the
end of the last subsection hold for what concerns modifications  of the
initial velocity variance.

\subsection{Energy of the fields}
\indent\noindent

In order to study the decoupling of the fields, let us expand the
effective potential of eq.(\ref{pot}) around the absolute minimum
${\hat\phi}^{\alpha}\equiv(v,{\vec 0})$.  The excitations of the fields are
$\mbox{$(\chi,{\vec\phi})=\phi^{\alpha}-{\hat\phi}^{\alpha}$}$
(thus $\chi$ is proportional to the fluctuations of the $\sigma$
field, whereas ${\vec\phi}\propto {\vec\pi}$ see eq.(\ref{adfield})).

Then, by a straightforward calculation, the quadratic part of the excitation
energy density at each site can be separated from the interaction term
\bqn E_{\chi}&=&\sum_{sites}~~\left\{
{{\dot{\chi}}^{2}\over 2}~+~ {\left(\nabla
  {\chi}\right)^2\over 2}~+~2\left(2|v|-1\right)\chi^2\right\}\nn\\
  E_{\phi_{i}}&=&\sum_{sites}~~\left\{
{{\dot{\phi_{i}}}^{2}\over 2}~+~ {\left(\nabla
    {\phi_{i}}\right)^2\over 2}~+~2\left(|v|-1\right)\phi^2\right\}
{}~~~~~~~~~~~i=2,3,4\nn\\
    E_{int}&=&\sum_{sites}~~
{4\over 3}\left\{\left[(v+\chi)^2+{\vec\phi}^2
    \right]^{3/2}-|v|^{3}\left[1~+~ 3{\chi\over v}~+~3{\chi^2\over
      v^2}~+~{3\over 2}{{\vec\phi}^2\over v^2} \right]\right\}
\label{energies}\eqn
The sum of these terms gives the total energy
$E_{tot}=\sum_{sites}~H$, where $H$ is the Hamiltonian defined
in eq.(\ref{ham}).

In Fig.\ref{energytime} we plot the evolution of each energy term,
normalized to the total energy, for
$\delta{\dot\phi}^{2}_{\alpha}=0.01~(a),~0.1~(b),~1~(c)$.  As already
mentioned, we notice that between $t\sim3fm$ and $t\sim7fm$ the pion
fields, and expecially the $\pi_2$, are amplified whilst the $\sigma$
energy gets reduced. Furthermore, at $t\sim10fm$ the interaction term
strongly reduces, and the fields practically decouple (even if a
small interaction term between $\sigma$ and $\pi$ is still present).
{}From the plots of Fig.\ref{energytime}, it is evident that increasing
the initial energy via the velocity variance leads to an increase of
the pions background, whereas it reduces the amplification factor of
the pions energy. Again this effect can be explained as due to the
action of the incoherent field fluctuations induced by the random
initial velocities which oppose themselves to the coherence of the DCC
phenomenon.  A quantitative computation of the mean value of the
energy of the $\pi_{2}$ field for times $t\simge10fm$, in
Fig.\ref{energytime}, gives
$ E_{\pi_{2}}\simeq215N~MeV~(a),~225N~MeV~(b),~440N~MeV~(c) $.
where $N$ is the number of fermion species.
We repeat, however,
that quantitative predictions are beyond the scope of this
simulation.  Furthermore the model, besides the limitation to be
bidimensional in the space coordinates, is characterized by an
undetermined number of fermion species $N$.  Of course it would be
interesting above all to evaluate the number of produced pions, which
in principle could be done for instance in the framework of coherent
states.

\subsection{Source dimension}
\indent\noindent

In this subsection we want to discuss the role of the size of the
source. We notice that it may be  plausible that there
exists a critical volume (a surface in our simulation) defining a lower
size for the source, below which no macroscopic
correlation phenomena should take place.
This can be qualitatively argued following
ref.\cite{rw}.  Let us consider the Fourier transform of the equation
of motion that we derive from the Hamiltonian (\ref{ham}), where we
have approximated the nonlinear term by its spatial average
$\langle|\phi|\rangle(\tau)$ \bqn \frac{d^2}{d\tau^2}
\tilde{\phi}_{\alpha}(\vec{k},\tau) = \left[-{\vec k}^2~+~4~-~
4\langle|\phi|\rangle(\tau)\right]\tilde{\phi}_{\alpha}(\vec{k},\tau)
{}~-~\lambda\delta_{\alpha 1}
\label{kmax}
\eqn
At the beginning of the expansion, when $\bra|\phi|\ket\simeq0$,
the enhanced modes are those with $ k < k_{max} \simeq 2$ (in dimensionless
units), which
correspond to a minimal wavelength of about $2fm$.
Anyway, if the linear dimensions of the source are less than some scale of
an order of magnitude determined by $k_{max}(t)$,
the growth of long-wavelength modes cannot take
place.  We have verified this for a source
of radius $0.5fm$: no correlated regions form.

\subsection{The case of annealing}
\indent\noindent

Finally, we show a sketch of
the time evolution of a different initial configuration in
Fig.\ref{annealing}. Here $\phi_{\alpha}$ are initially located
at the absolute minimum everywhere (annealing), whereas
an energetic
source (of $1~fm$ radius) is still present due to the spatial
modulation of the velocity variance. This is done
with an exponentially decreasing function whose
tail now extends over the whole lattice \cite{ahw}.
We show these pictures to stress that
the formation and growth of $R$-correlated regions can be present
without any $\theta$ disorientation, only coming from the spatial
distribution of the fluctuation energy. In this case the phenomenon of
growth proceeds by following a process analogous to spinodal
decomposition in alloys, since the equality of the areas comes out
to be approximately conserved in time.
Anyway this phenomenon occurs in regions where the energy density
is very low, and thus it is not relevant to produce correlated pions.
Furthermore it is washed out by the expansion of the source, for later times:
 this is due to the fact that the source carries out most
of the energy, which is not spatially ordered. In conclusion
the relevant features appear to be those related to the modeling
of the central part of the source, rather than to the other details,
such as the tail of the energy decay.

\newpage

\sect{Conclusions and Outlook}
\indent\noindent

In our simulation we have adopted initial conditions
which allow to control the energy of the system, and we have employed
fine lattice spacing.
Among the quantities we have computed there are
the $\theta$ angle between the $\sigma$ and $\pi$ fields,
the local ratios $R_{a}=\pi_{a}^2/{\vec\pi}^2$ $(a=1,2,3)$,
the quadratic term of the energies of the fields,
and the higher order term, for representative times.
Of course all of them are intimately related,
but there is no redundancy in the description
as some of them appear to be the more appropriate ones
according to the stage of the expansion.
Actually we remark that during the first time of
the expansion the $\sigma$ and $\vec\pi$ fields are strongly coupled
and thus the language in terms of asymptotic particles can be misleading.
On the contrary quantities related to the $O(4)$
angles are always well defined and the picture
emerging is clarified by the time evolution of the
whole set of the parameters.\hfb

The $\theta$ angle drives the formation of DCC,
whilst the crucial feature we look for is
the coherence of the pion field
oscillations in its internal $O(3)$ space, which is rather
signaled by the formation and growth of $R_{a}$-s
correlated regions.
Nevertheless these features
are linked, since from our numerical simulation it is clearly
confirmed  that the formation and growth of $R_{a}$-s correlated
regions is a macroscopic effect only when the $\theta$ angle
undergoes relevant and coherent disorientations during the expansion.
Actually the phenomenon of growth appears to be driven by the
energy excess at different points which
tends to align the pion field along a same direction.
Now, if the dynamics produces a strong and coherent $\theta$
fluctuation, a relevant part of the $\sigma$-field energy
is transferred to the pion field.
This happens in one of the equally-likely $O(3)$ directions
and generally manifests itself in a big
energy excess over long wavelengths, which is able to produce a macroscopic
effect on the growth phenomenon in terms of the $R_{a}$'s.
We recall that the $\sigma$
and $\vec\pi$ fields have to be nearly degenerate only in the central
part of the source, whereas most of the initial energy is
carried by the $\sigma$ field.\hfb

We also show that phenomena of growth can be present even for
very small $\theta$ disorientation, due to a non uniform spatial
energy distribution over a certain scale.
Anyway, in this case, the effect is much less important
and furthermore model-dependent.\hfb

For large enough times, when the interaction energy is small
(for instance, with a source of radius $1~fm$, it is enough
to have $t\simge10~fm$),
all the fields are almost decoupled, their energies are
separately conserved by the dynamics,
and one can talk of particles energy and number.

The indication we get from this study is that the relevant feature
to get large $R$-correlated regions (and thus
also to produce an unusual pion distribution), is
the initial energy release in the ``hot" vacuum and
its relative weight with respect to that at the
boundary (which is related to the dimension of the source and
to the shape of the energy decay out of the source).
We also find that, under these conditions,
an increase in the initial energy density leads generally
to a reduction of the relative energy
enhancement of the pion fields.\hfb

The nice feature resulting from this study is that we get a clear
description of the DCC mechanism by
means of $O(4)$-angles, and that we can allow for
a complete description of the final-states.
Although still not apt to give physical quantitative results because
of its low dimensionality, our work offers a guide to the exploration
of the driving features of the mechanism itself.
Among them, our simulation confirms that the role of random
fluctuations is to disturb the formation of large regions of
coherence, whereas it further suggests that the size of the ``hot"
source has to be large enough in order to produce
a macroscopic effect.
Following these indications, a large amount of energy release,
compatible with a large enough number of produced pions and with
small field fluctuations, can be obtained
only by increasing the source size.
This could hopefully occur if the expanding
fireball cools near $T_{c}$ having reached a linear
size of at least some $fm$.
Other details, such as the shape of the energy decay, look
less important.\vspace{2cm}

\centerline{\large\bf Acknowledgements}\hfb

We thank Michel Droz and Marco Pettini
for conversations on the subject of this work.

\newpage
\appendix{
\indent\noindent
We summarize briefly the algorithm used in the numerical
integrations. We have always used a second order algorithm, belonging to the
class of the so-called ``bilateral algorithm'' (see ref. \cite{case});
it has the remarkable property of being symplectic, i.e. it performs a
canonical transformation at every step of integration, thus faithfully
preserving the qualitative structure of the dynamics. The
simplecticity also ensures a very good energy conservation, while the
use of a second order algorithm allows for the use of comparatively long
integration steps (a second order algorithm is one for which the error
introduced at each step is ${\sf O}(\Delta t^3)$) without
precision-loss, hence reducing the cpu-time needed for the
simulation.  The explicit scheme of the algorithm used is the
following:
\begin{equation}
\left\{
\begin{array}{l}
\tilde{q}_i=q_i(t) \\[0.2cm]
\tilde{p}_i=p_i(t)+\frac{1}{2}\Delta t\,f_i[q(t)] \\[0.2cm]
q_i(t+\Delta t)=\tilde{q}_i+\Delta t\,\tilde{p}_i \\[0.2cm]
p_i(t+\Delta t)=\tilde{p}_i+\frac{1}{2}\Delta t\,f_i[q(t+\Delta t)] \\[0.2cm]
\hat{p}_i=p_i(t+\Delta t) \\[0.2cm]
\hat{q}_i=q_i(t+\Delta t)+\frac{1}{2}\Delta t\,\hat{p}_i \\[0.2cm]
p_i(t+2\Delta t)=\hat{p}_i+\Delta t\,f_i(\hat{q}_i) \\[0.2cm]
q_i(t+2\Delta t)=\hat{q}_i+\frac{1}{2}\Delta t\,p_i(t+2\Delta t). \\[0.2cm]
\end{array} \right.
\label{symp2s}
\end{equation}
(Here $f_{i}$ are the forces.
$\tilde{q}_i,\tilde{p}_i=p_i,\hat{q}_i,\hat{p}_i$ are dummy
variables lacking of any physical meaning).
The scheme is called bilateral because it performs two integration
steps at a time by the r\^ole of the $q$ and $p$. This
feature results in an enhancement of the precision with which energy
is conserved. In typical simulations, we found that energy fluctuated
around its mean value with a relative amplitude of about $10^{-5}$
having set the time step to $0.01$. It is also remarkable that the
mean value of these fluctuations vanishes: this ensures that the
system is not subject to spurious damping or excitations.
(For further details of the method see ref.\cite{case}).
}

\newpage
\centerline{\Large\bf Figures}

\begin{list}{\bf Fig.\theenumi}{\usecounter{enumi}}
\item{Plot of the effective potential
of eq.(\protect\ref{pot}) vs. $\phi_{1}\propto\sigma$,
at ${\vec\pi}=0$ (the vacuum energy density has been
subtracted).
\label{potential}
}

\item{Plot of the $\phi_{1}\propto\sigma$ configuration on the lattice
at $t=0.02 fm$ ($\delta{\dot\phi}^2_{\alpha}=0.1$).
\label{initialsigma}
}

\item{Plot of the $\phi_{3}\propto\pi_2$ configuration on the lattice
at $t=0.02 fm$ ($\delta{\dot\phi}^2_{\alpha}=0.1$).
\label{initialpi}
}

\item{Plot of the energy density shape at
 $t=0.02fm$, on the internal $50\protect\times50$ lattice centered in
 the source (see eq.(\ref{ham}).
 Here $\delta{\dot\phi}^2_{\alpha}=0.1$. \label{initialen}
}

\item{Plots of the $\theta$ angle on the internal
$32\protect\times32$
square lattice centered in the source (initially the
source radius is $\rho\simeq10$ sites, Fig.$(a)$),
 for representative times ($\delta{\dot\phi}^2_{\alpha}=0.1$).
 The range $[0,\pi]$  has been  divided in
 five  equally
spaced regions colored according to a gray scale,
 ranging from black ($\theta\simeq0$) to white ($\theta\simeq\pi$).
These figures explicitly show three different behaviours. At
early times there is a formation and growth of $\theta $ correlated
regions inside the source. Such regions tend to melt together in
the bulk, while the outer regions align to $\theta=\pi $.
This phenomenon ends up at about $t=1fm$, when the whole source
is aligned with the true vacuum. Anyway, the system has not relaxed to
the true vacuum yet. From the center of the source, in fact, a collective
motion starts, which brings, at $t\simeq2.5fm$, the whole source
to be at $\theta=0$.
Afterwards, the system begins to perform damped oscillations
along the bottom of the
potential well.
\label{thetaregions}
}

\item{In Fig.$(a)$ we plot $\langle\theta (r)\rangle (t)$. The data refer
to the same simulation as in Fig.\protect\ref{thetaregions}. The mean
value is taken on concentric rings of thickness $0.4 fm$.
This picture points out the time scales
characterizing the different stages of the phenomenon of DCC.
In Fig.$(b)$ we plot the shape of two sections at fixed radius ${\bar r}$,
with $0 <{\bar{r}}<0.4~fm$ (continuous) and
$0.8<{\bar{r}}<1.2~fm$ (dashed).
\label{thetaslope}
}

\item{In these figures the  $\theta$ configurations on the $(x,y)$
lattice is given by  the arrow orientations.
These pictures are taken in coincidence of some maxima of
disorientation as it can be seen in Fig.\ref{thetaslope}
(they refer to the same simulation).
As in Fig.\ref{thetaregions}, it can be noticed that
the disorientation is spatially  incoherent
at $t=0.02fm$ (Fig.$(a)$), whereas
it has become coherent at $t=2fm$ and $t=5fm$
(Fig.$(b)$-$(c)$).
Finally, in Fig.$(d)$, at $t=9fm$,
the presence of concentric regions of different $\theta$
orientation is evident. To get a satisfying resolution in Fig.$(a)$ and $(b)$
only the central $25\times 25$ lattice has been plotted (outside $\theta=\pi$),
whereas the central $50\times 50$ in Fig. $(c)$ and
$100\times100$ lattice in Fig.$(d)$ have been taken into account. Then
the average over $2\times 2$ (Fig.$(c)$) and $4\times4$ blocks
(Fig.$(c)$) has been performed.
\label{vectorfields}
}

\item{
Distributions of
the representative points of the lattice sites in the $(\varphi,\eta)$
plane (see eq.(\protect\ref{fiesta})). The data refer to the same simulation
as in the previous pictures. At the beginning of the
simulation all the directions are
equally likely and thus the distribution is uniform.
Then the existence of preferred directions in which the
pion field oscillates starts to be evident
(from $t\simeq 4fm$) and becomes more evident for later times.
The residual noise which is still visible
for $t=12 fm$ in Fig.$(d)$ is mostly due to the points of the lattice
which are not yet reached from the expanding source.
\label{etaphi}
}

\item{Configuration of the ratio $R_{2}$
($R_{a}\equiv \pi_{a}^2/{\vec\pi}^2$) at each lattice site
for representative
 times ($\delta{\dot\phi}^2_{\alpha}=0.1$).
 To get a pictorial representation, the interval of definition of $R_{a}$
has been divided in three regions which are initially
equally probable: black if $R_{a}\in [0,1/9]$,
grey if $R_{a}\in [1/9,4/9]$ and white if $R_{a}\in [4/9,1]$.
These pictures show the formation of large $R_{2}$-correlated regions in which
the pion field oscillates along a preferred direction
(see also Fig.\protect\ref{etaphi}).
\label{R2}
}

\item{ Plots of the quadratic terms of the energy
of each field and of the higher order term
(see. eq.(\protect\ref{energies})), normalized to
the total energy, vs. time. With reference to
eq.(\protect\ref{energies}), the curves give the behaviour
respectively of: $E_{\chi}$ (continuous),
$E_{\phi_{2}}$ (dotted-dashed), $E_{\phi_{3}}$ (dashed),
$E_{\phi_{4}}$ (dotted-long dashed), $E_{int}$ (dotted).
We remember that
$(\phi_{2},\phi_{3},\phi_{4}) \propto (\pi_{1},\pi_{2},\pi_{3})$.

 Notice that  between $t\simeq3fm$ and $t\simeq7fm$
 the pion fields, and expecially
the $\pi_2$ field, are amplified.
Furthermore, at $t=10fm$ the interaction
term strongly reduces, and each field behave almost as free.
It is also evident
that the increase of the initial velocity variance
(from Fig.$(a)$ to Fig.$(c)$)
leads to an increase of the pions background, whereas it reduces
their energy amplification.
\label{energytime}
}

\item{Plot of the ratio $R_2$ in the case in which
the system is initially set at the bottom of the
potential well of Fig.\protect\ref{potential}.
 The initial velocity variance has been modulated with
 an exponentially decreasing function whose tail extends over the whole
 lattice \protect\cite{ahw}. These figures show
 that the formation and growth of
 $R$-correlated regions can be present even without any $\theta$
 disorientation, due only to the spatial distribution of the energy.
 Anyway this phenomenon occurs in regions where the energy density
 is very low, and thus it is not relevant to produce correlated pions.
 Furthermore it is washed out by the expansion of the source, for later times.
\label{annealing}
}
\end{list}

\end{document}